\documentclass[%
 reprint,
 nofootinbib,
 amsmath,amssymb,
 aps,
 prl,
]{revtex4-1}

\usepackage{graphicx}
\usepackage{dcolumn}
\usepackage{bm}
\usepackage[colorlinks,linkcolor=black,anchorcolor=blue,citecolor=blue,
urlcolor=blue]{hyperref}
\usepackage{multirow}

\newcommand{\pauli}[1]{\sigma_{\mathrm #1}}
\newcommand{\ii}{\mathrm{i}}
\newcommand{\anni}[1]{a_{#1}}
\newcommand{\crea}[1]{a_{#1}^{\dagger}}
\newcommand{\affi}{\affiliation{Center for Quantum Information, Institute for Interdisciplinary Information Sciences, Tsinghua University, Beijing, China}}
\newcommand{\yb}{^{171}\mathrm{Yb}^{+}}
\newcommand{\diff}[1]{\mathrm{d} #1}

\newcommand{\ket}[1]{\left| #1 \rangle \right.}
\newcommand{\incite}[1]{~\cite{#1}}
\newcommand{\oncite}[1]{Ref.~\cite{#1}}
\newcommand{\figref}[1]{Fig.~\ref{#1}}
\newcommand{\equref}[1]{equation~(\ref{#1})}
\newcommand{\tabref}[1]{Table~\ref{#1}}

\begin{document}

\title{
Scalable global entangling gates on arbitrary ion qubits
}

\author{Yao \surname{Lu}}
 \email{luyao\_physics@163.com}
 \affi
\author{Shuaining \surname{Zhang}}
 \affi
\author{Kuan \surname{Zhang}}
 \affi
\author{Wentao \surname{Chen}}
 \affi
\author{Yangchao \surname{Shen}}
 \affi
\author{Jialiang \surname{Zhang}}
 \affi
\author{\\Jing-Ning \surname{Zhang}}
 \affi
\author{Kihwan \surname{Kim}}
 \email{kimkihwan@mail.tsinghua.edu.cn}
 \affi

\date{\today}

\begin{abstract}
A quantum algorithm can be decomposed into a sequence consisting of single qubit and 2-qubit entangling gates. To optimize the decomposition and achieve more efficient construction of the quantum circuit, we can replace multiple 2-qubit gates with a single global entangling gate. Here, we propose and implement a scalable scheme to realize the global entangling gates on multiple $\yb$ ion qubits by coupling to multiple motional modes through external fields. Such global gates require simultaneously decoupling of multiple motional modes and balancing of the coupling strengths for all the qubit-pairs at the gate time.
To satisfy the complicated requirements, we develop a trapped-ion system with fully-independent control capability on each ion, and experimentally realize the global entangling gates.
As examples, we utilize them to prepare the Greenberger-Horne-Zeilinger (GHZ) states in a single entangling operation, and successfully show the genuine multi-partite entanglements up to four qubits with the state fidelities over $93.4\%$.
\end{abstract}

\maketitle

\begin{figure*}[htb]
  \centering
  \includegraphics[scale = 0.75]{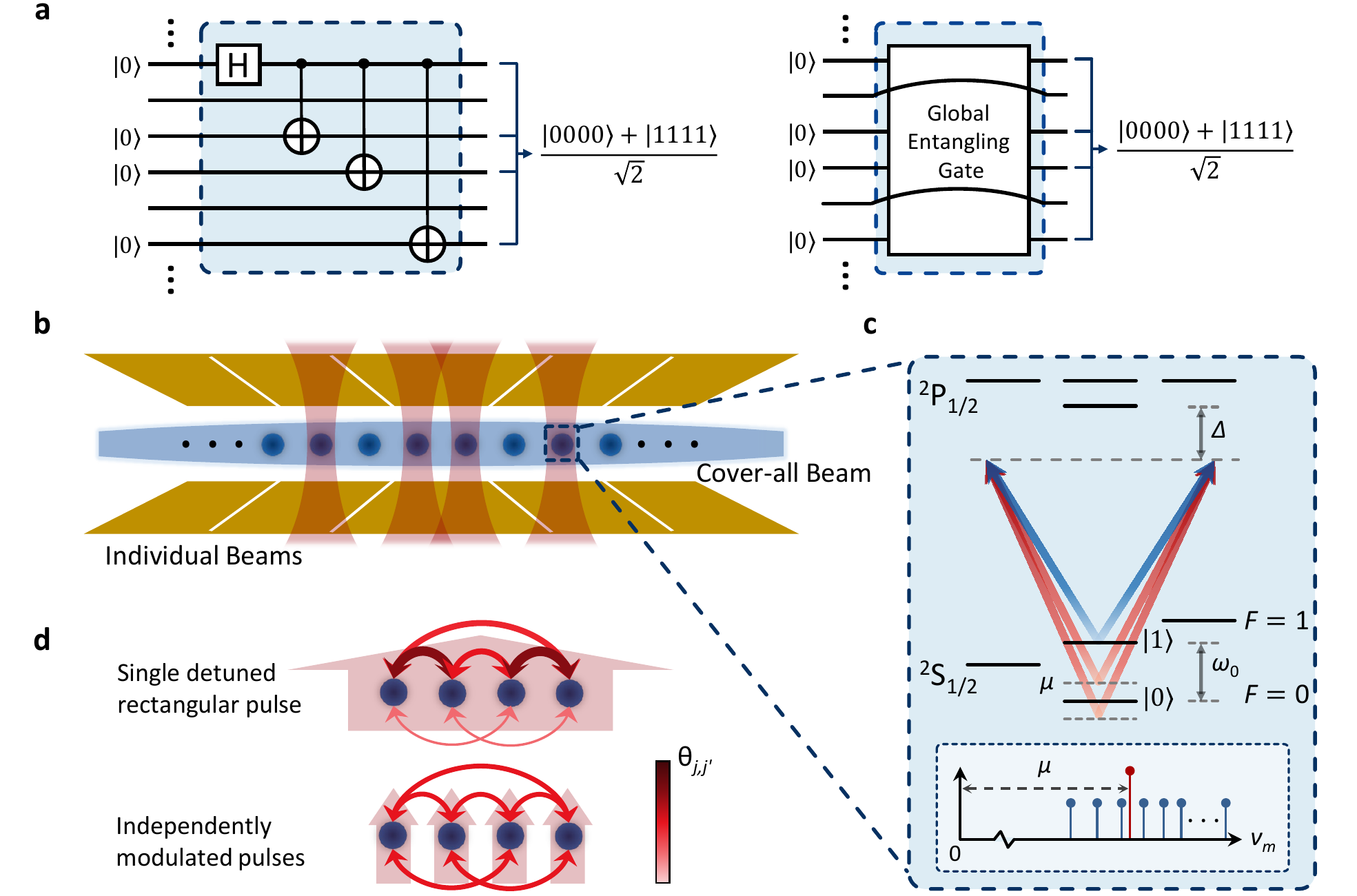}
  \caption{
  {\bf Global entangling gate and its experimental implementation.}
  {\bf a.} Efficient quantum-circuit construction by using global gate. For the generation of the $N$-qubit GHZ state, we need $N-1$ times of pairwise entangling gates, which can be replaced by a single global entangling gate up to single qubit rotations. Here we give an example of 4-qubit GHZ state generation among arbitrary number of qubits.
  {\bf b.} Experimental setup for the implementation of the global entangling gate. Each ion in the trap encodes a qubit, which is individually manipulated by Raman beams that consist of a cover-all beam (blue) and a single-ion addressing beam (red). The individually addressed qubits are involved in the global entangling gate.
  {\bf c.} Energy levels of $\yb$ and motional structure. The Raman beams introduce qubit-state dependent force on each ion, with the driving frequency $\mu$, which drives the multiple motional modes simultaneously.
  {\bf d.} Implementation of the global entangling gate. With a single constant pulse we cannot achieve uniform coupling on all the qubit-pairs due to the lack of enough controllable parameters. Instead, we can have uniform coupling by independently modulating the pulses on each ion.
  }\label{fig:setup}
\end{figure*}

Quantum computers open up new possibilities of efficiently solving certain classically intractable problems, ranging from large number factorization\incite{shor1997polynomial-time} to simulations of quantum many-body systems\incite{Feynman1982Simulating,lloyd1996universal,Blatt2012Quantum}. Universal quantum computation tasks, e.g. quantum phase estimation\incite{Nielsen2010Quantum}, Shor's algorithm\incite{shor1997polynomial-time,Monz2016Realization} and quantum variational eigensolver\incite{peruzzo2014a,shen2017quantum}, can be decomposed by single qubit and 2-qubit entangling gates in the quantum circuit model\incite{Nielsen2010Quantum}. However, such decompositions are not necessarily efficient\incite{Ivanov2015Efficient,martinez2016compiling,Maslov2018Use}. Recent theoretical works have pointed out that, with the help of global $N$-qubit entangling gates ($N>2$), it is possible to have the polynomial or even exponential speed up in constructing various many-body interactions\incite{Casanova2012Quantum,yung2015from,garciaalvarez2017digital,Hempel2018Quantum} and build more efficient quantum circuits for innumerous quantum algorithms\incite{Ivanov2015Efficient,martinez2016compiling,Maslov2018Use}. For example, the $N-1$ pairwise entangling operations in the preparation of the $N$-qubit GHZ state can be replaced by a single global entangling gate, shown in \figref{fig:setup}~(a).

The global entangling gates demand fully-connected couplings among all the qubits, which naturally emerge in trapped-ion systems\incite{Kim2009Entanglement,korenblit2012quantum,linke2017experimental}. The ion qubits are entangled by coupling to the collective motional modes through external fields\incite{molmer1999multiparticle,sorensen1999quantum,solano1999deterministic}, leading to the all-to-all network. Previously, the global entangling gates have been realized by only coupling to the axial center-of-mass (COM) mode\incite{lanyon2011universal,barreiro2011an,Monz201114}. However, the single mode approach is hard to scale up because the isolation of the mode is challenging as the number of ions increases in a single crystal\incite{Monz201114,Zhu2006Arbitrary,zhu2006trapped}. Recently, a scalable scheme, by driving multiple motional modes simultaneously, have been proposed to achieve 2-qubit gates with modulated external fields\incite{Zhu2006Arbitrary,zhu2006trapped,Steane2014Pulsed,Green2014Phase,Leung2018Robust} and already been demonstrated in the experiments\incite{Choi2014Optimal,Leung2018Robust,milne2018phase}. However, no one has explored the possibility of applying this multi-mode scheme to the global $N$-qubit case yet, either theoretically or experimentally.

Beyond the 2-qubit gate, for the first time, we develop and demonstrate the global multi-qubit entangling gates by simultaneously driving multiple motional modes with modulated external fields in a fully-controllable trapped-ion system. Compared with the 2-qubit situation, we not only need to decouple the qubits from all the motional modes simultaneously at the gate time, but also have to satisfy more constraints coming from the coupling strengths of all qubit pairs. We derive the theoretical expressions of all the constraints and find out it is possible to construct the global entangling gate with the modulated fields. In order to fulfill all the theoretical requirements, we establish the trapped-ion system with the capability of independent control of the parameters of the external fields on each qubit, as shown in \figref{fig:setup}~(b). As a proof of principle demonstration, we realize the global entangling gates up to four $\yb$ ions, which we use to create the GHZ states. Moreover, we show the global entangling gate works on an arbitrary subset of the entire ion-chain, by simply turning off the external fields on the ions out of the subset.

In the experiment, we implement the global entangling gate in a single linear chain of $\yb$ ions. A single qubit is encoded in the hyperfine levels belonging to the ground manifold $^2S_{1/2}$, denoted as $\ket{0} \equiv \ket{F = 0, m_F = 0}$ and $\ket{1} \equiv \ket{F = 1, m_F = 0}$ with the energy gap $\omega_0 = 2\pi \times 12.642821~\mathrm{GHz}$\incite{fisk1997accurate}, as shown in \figref{fig:setup}~(c). The qubits are initialized to the state $\ket{0}$ by the optical pumping and measured by the state-dependent fluorescence detection\incite{Olmschenk2007Manipulation}. The fluorescence is collected by an  electron-multiplying charge-coupled device (EMCCD) to realize the site-resolved measurement. Additional information about the experimental setup is shown in the Methods section.

After ground state cooling of the motional modes, the coherent manipulations of the qubits are performed by Raman beams produced by a pico-second mode-locked laser\incite{hayes2010entanglement}. One of the Raman beams is broadened to cover all the ions, while the other is divided into several paths which are tightly focused on each ion. The cover-all beam and the individual beams perpendicularly cross at the ion-chain and drive transverse modes mainly along x-direction. With the help of the multi-channel acousto-optic modulator (AOM) controlled by the multi-channel arbitrary waveform generator (AWG), we realize independent control of the individual beams on each ion, as illustrated in \figref{fig:setup}~(b), which is similar to \oncite{Debnath2016Demonstration}.

To perform the global entangling gates with the form of
\begin{equation}\label{equ:UGEG}
  \exp
    \left[
      - \ii~
      \dfrac{\pi}{4}
      \sum_{j<j'}^{N}
      \pauli{x}^{j}\pauli{x}^{j'}
    \right],
\end{equation}
we apply the modulated bichromatic Raman beams with the beat-note frequencies $\omega_0 \pm \mu$ to the whole ion-chain, where $\mu$ is the detuning from the carrier transition and its value is around the frequencies of the motional modes. The above bichromatic beams lead to the qubit state-dependent forces on each qubit site\incite{haljan2005spin-dependent,Lee2005Phase} and the time evolution operator at the gate time $\tau$ can be written as\incite{Zhu2006Arbitrary}
\begin{equation}\label{equ:evo_operator}
  U(\tau) =
    \exp
      \left[
        \sum_{j,m}
        \beta_{j,m}(\tau) \pauli{x}^{j}
    \\ - \ii \sum_{j<j'} \theta_{j,j'}(\tau) \pauli{x}^{j} \pauli{x}^{j'}
  \right].
\end{equation}
Here $\theta_{j,j'}(\tau)$ is the coupling strength between the $j$-th and the $j'$-th qubit in the form of
\begin{align}\label{equ:coupling}
  \nonumber
  \theta_{j,j'}(\tau) &
    = - \sum_{m}
    \int_{0}^{\tau} \diff{t_2}
    \int_{0}^{t_2} \diff{t_1}
    \dfrac{
        \eta_{j,m} \eta_{j',m}
        \Omega_j(t_2) \Omega_{j'}(t_1)}{2} \\
  &\sin\left[ (\nu_m - \mu)(t_2 - t_1) - \left( \phi_j(t_2) - \phi_{j'}(t_1) \right) \right],
\end{align}
where $\eta_{j,m}$ is the scaled Lamb-Dicke parameter\incite{James1998Quantum}, $\anni{m}$ ($\crea{m}$) is the annihilation (creation) operator of the $m$th motional mode, $\nu_m$ is the corresponding mode frequency, $\Omega_j(t)$ and $\phi_{j}(t)$ are the amplitude and the phase of the time-dependent carrier Rabi frequency on $j$-th ion. And $\beta_{j,m}(\tau) = \alpha_{j,m}(\tau) \crea{m} - \alpha_{j,m}^\ast(\tau) \anni{m}$, where $\alpha_{j,m}$ represents the displacement of the $m$-th motional mode of the $j$-th ion in the phase space, written as
\begin{equation}\label{equ:trajectory}
  \alpha_{j,m}(\tau) =
    - \ii \eta_{j,m}
        \int_{0}^{\tau}
        \dfrac{\Omega_j(t) e^{-\ii \phi_{j}(t)}}{2}
        e^{\ii \left( \nu_m - \mu \right) t}
        \diff{t}.
\end{equation}
Due to the interference of the multiple motional modes, it is not nature to have uniform coupling strengths on all the qubit pairs with single detuned rectangular pulse in a conventional manner, as shown in \figref{fig:setup}~(d). Instead, we can employ individual control of time-dependent parameters $\{ \Omega_j(t)$, $\phi_j(t) \}$, to satisfy the below constraints
\begin{align}
  \alpha_{j,m}(\tau) &= 0, \label{equ:condition_1} \\
  \theta_{j,j'}(\tau) &= \pi/4, \label{equ:condition_2}
\end{align}
for any motional modes $m$ and qubit pairs $j,j'$. We note that, once we find the solution of the global $N$-qubit entangling gate, the entangling gate on any subset qubits can be straightforwardly applied by simply setting $\Omega_{j} = 0$ for the qubit $j$ outside the subset.

Considering a system with $N$ qubits and $M$ collective motional modes, there are $N\times M$ constraints from the requirements of the closed motional trajectories and $\binom{N}{2}$ from the conditions of the coupling strength. Therefore we have to satisfy a total number of $N(N-1)/2 + NM$ constraints. In principle, we can fulfill the constraints by modulating the intensities and the phases of the individual laser beams continuously or discretely. In the experimental implementation, we choose discrete phase modulation because we have high precision controllability on the phase degree of freedom. We divide the total gate operation into $K$ segments with equal duration and change the phase on each ion in each segment, which provides $N\times K$ independent variables. Because of the nonlinearity of the constraints, it is challenging to find analytical solutions of the constraints \equref{equ:condition_1} and \equref{equ:condition_2}. Therefore, we construct an optimization problem to find numerical solutions. We minimize the objective function of $\sum_{j,m}\left|\alpha_{j,m}(\tau)\right|^2$\incite{hayes2012coherent,Leung2018Robust,webb2018resilient,shapira2018robust} subject to the constraints of \equref{equ:condition_2}. Note that we also employ the amplitude shaping at the beginning and the end of the operation to minimize fast oscillating terms due to the off-resonate coupling to the carrier transition\incite{Roos2008Ion}.
The details of the constraints under discrete phase modulation and the optimization problem construction are provided in the Methods section.

To experimentally test the performance of the global $N$-qubit entangling gate, we use it to generate the $N$-qubit GHZ state and then measure the state fidelity. Starting from the product state $\ket{0\cdots0}$, the GHZ state can be prepared by applying the global entangling gate, while additional single qubit $\pauli{x}$-rotations by $\pi/2$ are needed if $N$ is odd. After the state preparation, we obtain the state fidelity by measuring the population of the entangled state and the contrast of the parity oscillation\incite{sackett2000experimental}. We also use the fidelity of the GHZ state to test the important feature of the global entangling gate, which is that we can realize entangling gates on any subset of qubits that are addressed by individual laser beams without changing any modulation pattern.

\begin{figure}[tb]
  \centering
  \includegraphics[scale = 0.74]{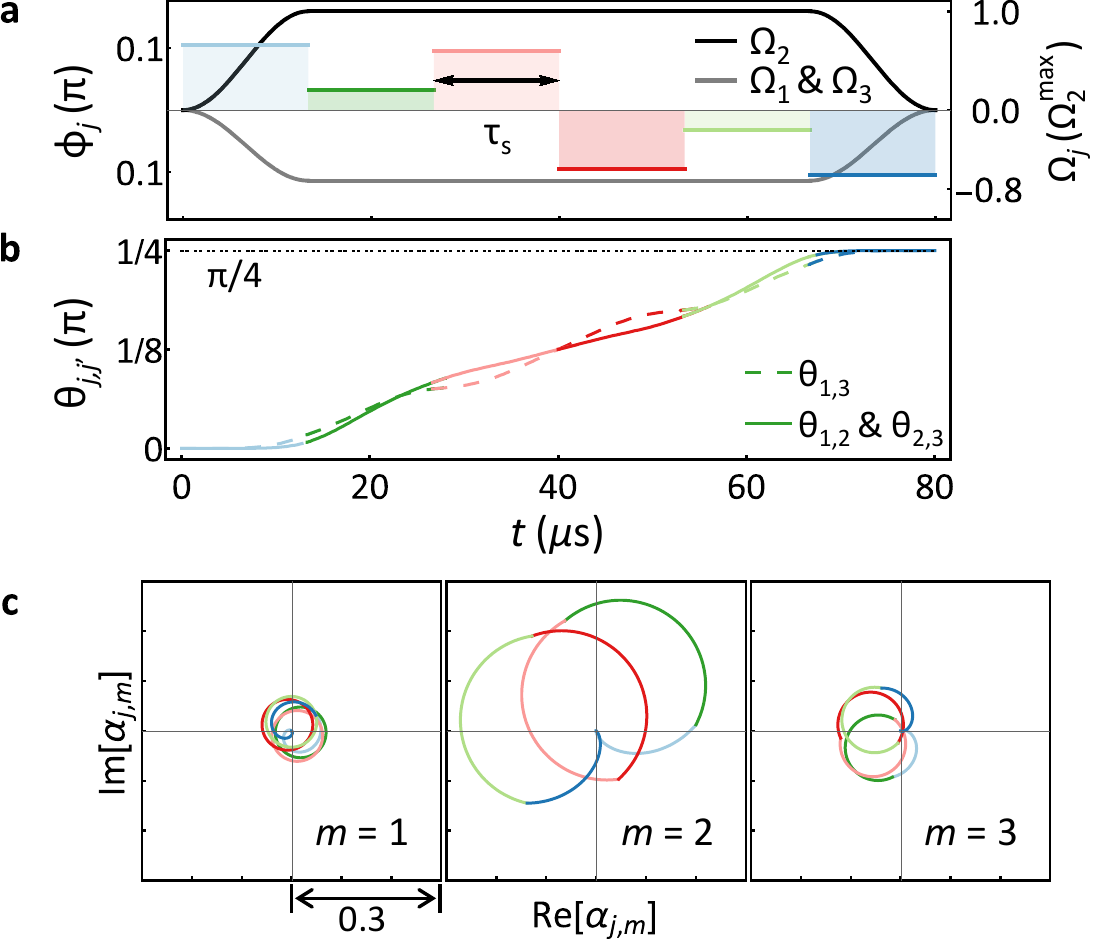}
  \caption{
  {\bf Experimental implementation of the global 3-qubit entangling gate.}
  {\bf a.} Pulse scheme with the phase and amplitude modulation. The phase $\phi_{j}(t)$ is discretely modulated, shown in the colored lines, which is same for all three ions. The specific values of the modulated phases are given in the Methods section. The Rabi frequencies, shown in the black and gray curves, are shaped at the beginning(end) of the gate operation using $\sin^2$-profile with the switching time equal to the duration of a single segment.
  {\bf b.} Accumulation of the coupling strengths over the evolution time. All the coupling strengths increase to the desired value of $\pi/4$.
  {\bf c.} Motional trajectories $\alpha_{j,m}$ for the 1st qubit in the phase space as an example. The different colors are corresponding to the different segments in {\bf a}.}\label{fig:3qubit_setting}
\end{figure}

\begin{figure}[htb]
  \centering
  \includegraphics[scale = 0.76]{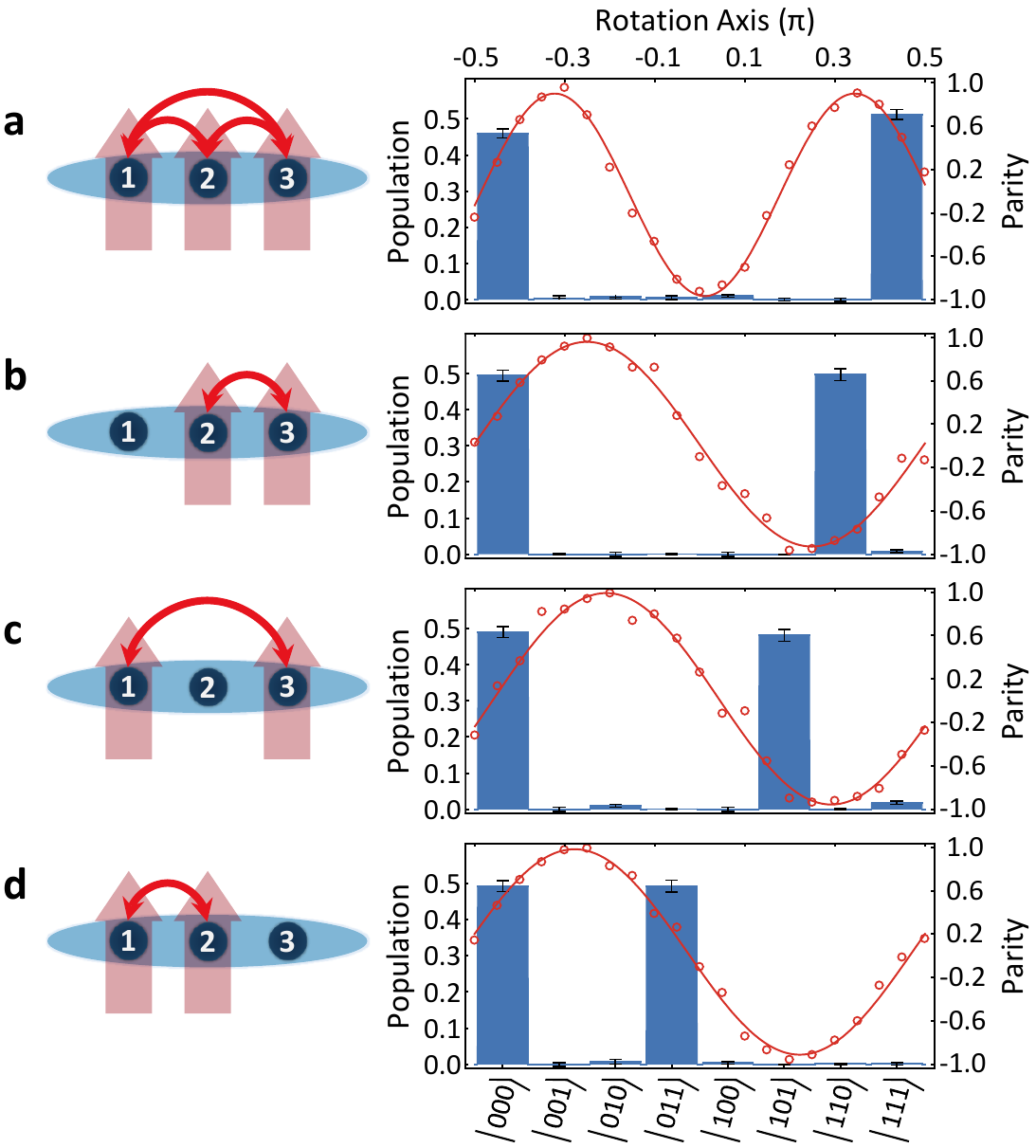}
  \caption{
  {\bf Experimental results of the global entangling gates in three ion-qubits.}
  Left column shows the capability of the global entangling gate, which can generate entanglement of {\bf a.} the entire qubits or {\bf b-d.} any pairs by switching on the individual beams on the target ions without changing any modulated patterns.
  Right column shows the population (blue histogram) and the parity oscillation (red circles for the experimental data and red curves for the fitting results) of the generated GHZ state.
  {\bf a.} 3-qubit GHZ state with the state fidelity of $\left( 95.2\pm1.5 \right)\%$.
  {\bf b-d.} 2-qubit GHZ states of qubit pairs $\left\{ 2,3 \right\}$, $\left\{ 1,3 \right\}$ and $\left\{ 1,2 \right\}$, with the fidelities of $\left( 96.7 \pm 1.8 \right)\%$,$\left( 97.1 \pm 1.9 \right)\%$ and $\left( 96.5 \pm 1.5 \right)\%$, respectively.
  } \label{fig:3qubit}
\end{figure}

As the first demonstration, we use three $\yb$ ions with the frequencies of the collective motional modes in the x-direction $\left\{ \nu_{\mathrm{1}}, \nu_{\mathrm{2}}, \nu_{\mathrm{3}} \right\} = 2\pi \times \left\{ 2.184, 2.127, 2.044 \right\}~\mathrm{MHz}$. We choose the detuning $\mu$ to be $2\pi\times2.094~\mathrm{MHz}$, between the last two modes. The total gate time is fixed to be $80~\mathrm{\mu s}$ and divided into six segments. The details of the phase modulation pattern and the amplitude shaping with relative ratio are shown in \figref{fig:3qubit_setting}~(a). With these parameters, the constraints of \equref{equ:condition_1} and \equref{equ:condition_2} are fulfilled, shown in \figref{fig:3qubit_setting}~(b-c). We use this global 3-qubit entangling gate to prepare the 3-qubit GHZ state with the state fidelity of $\left( 95.2\pm1.5 \right)\%$, as shown in \figref{fig:3qubit} (a).

Moreover, by turning off the individual beam on a qubit, we can remove the couplings between the qubit and the others, as shown in \figref{fig:3qubit}. In the 3-qubit system, the global entangling gates on the subsets become the pairwise gates on the arbitrary qubit-pair, which are used to generate the 2-qubit GHZ states with the fidelities over $96.5\%$ in the experiment, as shown in \figref{fig:3qubit} (b-c).

\begin{figure*}[htb]
  \centering
  \includegraphics[scale = 0.67]{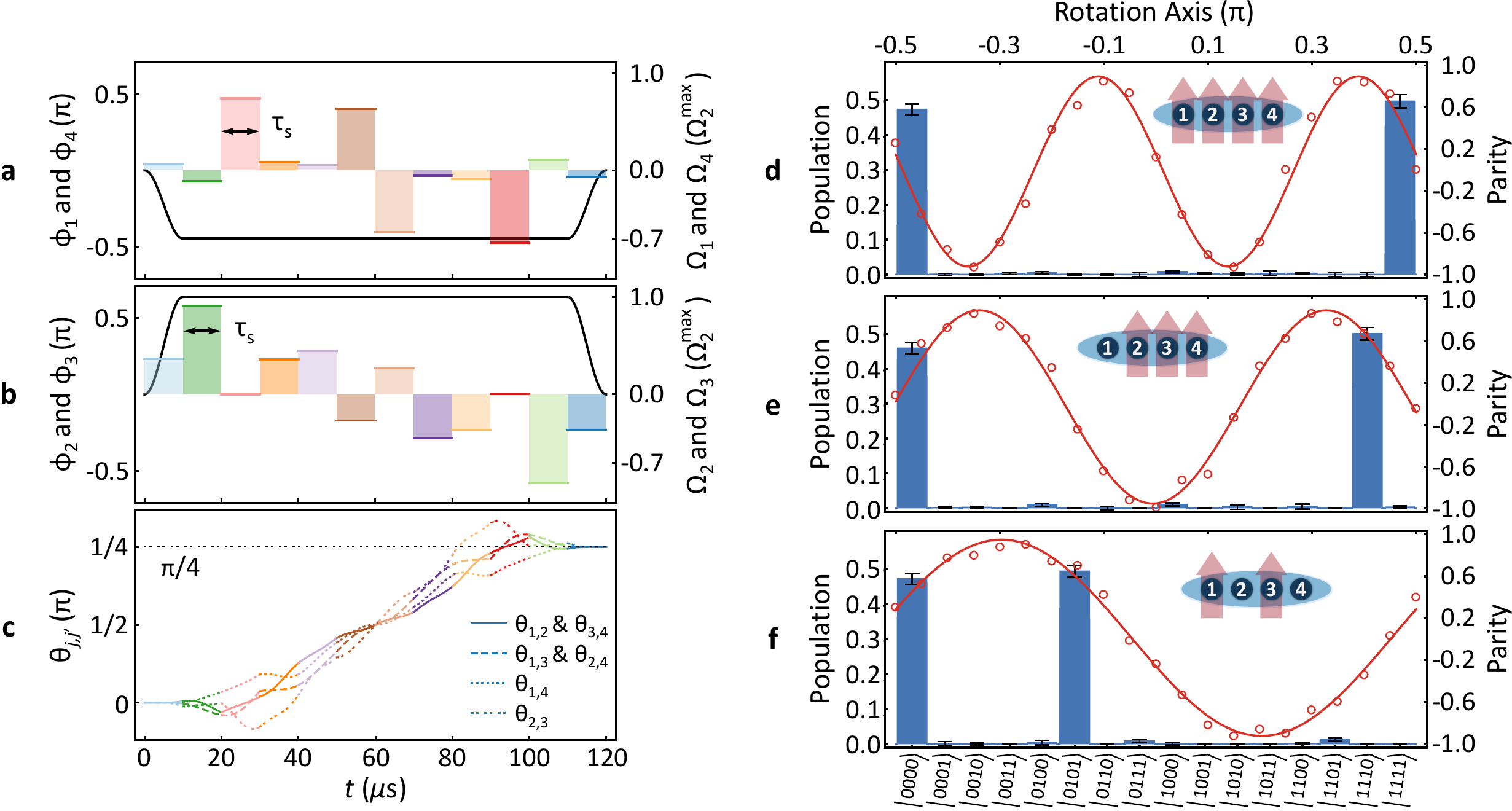}
  \caption{
  {\bf Experimental implementation and results of the global entangling gate in a four-ion system.}
  {\bf a-b.} Pulse scheme with the phase and amplitude modulation. Using the symmetry of the system, we set the modulation patterns to be same for the outer two qubits of $\left\{ 1,4 \right\}$ and the inner two of $\left\{ 2,3 \right\}$. The specific values of the modulated phases and the motional trajectories under this pulse scheme are shown in the Methods section.
  {\bf c.} Accumulation of the coupling strengths for all the qubit-pairs, which converge to the desired value of $\pi/4$ at the end of the gate.
  {\bf d-f.} The GHZ states prepared by the global entangling gates. By addressing an arbitrary subset of the qubits, for example $\left\{ 1,2,3,4 \right\}$, $\left\{ 2,3,4 \right\}$ and $\left\{ 1,3 \right\}$, we can perform the entangling gate on the subset. The frequency of the parity oscillation, proportional to the number of the addressed qubits, reveals the prepared state is the GHZ state. The state fidelities of the prepared 4-qubit, 3-qubit and 2-qubit GHZ states reach $\left( 93.4\pm2.0 \right)\%$, $\left( 94.2\pm1.8 \right)\%$ and $\left( 95.1\pm1.6 \right)\%$, respectively.
  }\label{fig:4qubit_gate}
\end{figure*}

For the demonstration of the scalability, we move to a 4-qubit system with the motional frequencies $\left\{ \nu_{1}, \nu_{2}, \nu_{3}, \nu_{4} \right\} = 2\pi \times \left\{ 2.186, 2.147, 2.091, 2.020 \right\}~\mathrm{MHz}$. The larger system means more constraints and more segments are required. To realize the global 4-qubit entangling gate, we choose the detuning $\mu$ to be $2\pi \times 2.104~\mathrm{MHz}$ and fix the total gate time to be $120~\mathrm{\mu s}$, which is evenly divided into twelve segments. The pulse scheme is shown in the \figref{fig:4qubit_gate} (a-b). The number of the constraints in \equref{equ:condition_2} increases quadratically with the number of the qubits and reaches to six in the 4-qubit case, as shown in \figref{fig:4qubit_gate} (c).

By applying the global 4-qubit entangling gate to all the qubits, we successfully generate the 4-qubit GHZ state with the state fidelity of $\left( 93.4\pm2.0 \right)\%$, as shown in \figref{fig:4qubit_gate}~(d). Similarly we can prepare the 3-qubit GHZ state or the 2-qubit GHZ state by only addressing arbitrary three or two qubits, respectively. Experimentally we choose the qubits of $\left\{ 2,3,4 \right\}$ to prepare the 3-qubit GHZ state and the qubit-pair of $\left\{ 1,3 \right\}$ to prepare the 2-qubit GHZ state, with the state fidelities of $\left( 94.2\pm1.8 \right)\%$ and $\left( 95.1\pm1.6 \right)\%$, respectively, as shown in \figref{fig:4qubit_gate}~(e-f).

All of the results are calibrated to remove the detection errors by using the method described in \oncite{Duan2012Correcting}. The state fidelities of all the prepared GHZ states are mainly limited by the fluctuations of the tightly focused individual beams and the optical paths jittering of the Raman beams ($2 \sim 4 \%$). Other infidelity sources in the experiment include the drifting of the motional frequencies ($1\sim2 \%$) and the crosstalk of the individual beams to the nearby ions ($\sim 1\%$).

We present the experimental realization of the global entangling gate, which can make quantum circuit efficient, in a scalable approach on the trapped-ion platform. Moreover, we theoretically optimize the pulse schemes for the five and six qubits and we find the required number of segments and the gate duration increase linearly with the number of qubits. So far we have not found the limitation to scale up the global entangling gate to a further number of qubits. However, the optimization of the pulse schemes with large number of qubits belongs to NP-hard problems, but could be assisted by classical machine learning technique. Furthermore, we can extend the global entangling gate to a general form with arbitrary coupling strengths of $\left\{\theta_{j,j}(\tau) = \Theta_{j,j'} \right\}$, which would provide further simplification of quantum circuits for large-scale quantum computation and simulation\incite{Maslov2018Use}. During the preparation of the paper, we have been aware of the related work about the parallel pairwise entangling gate\incite{figgatt2018parallel}.

\section*{Acknowledgements}
This work was supported by the National Key Research and Development Program of China under Grants No. 2016YFA0301900 (No. 2016YFA0301901) and the National Natural Science Foundation of China 11574002, and 11504197.

\subsection*{Author information}
These authors contributed equally: Yao Lu, Shuaining Zhang and Kuan Zhang. 



\begin{thebibliography}{10}

\bibitem{shor1997polynomial-time}
Peter~W Shor.
\newblock Polynomial-time algorithms for prime factorization and discrete
  logarithms on a quantum computer.
\newblock {\em SIAM J. Comput.}, 26(5):1484--1509, 1997.

\bibitem{Feynman1982Simulating}
Richard~P Feynman.
\newblock Simulating physics with computers.
\newblock {\em Int. J. Theor. Phys.}, 21(6-7):467--488, 1982.

\bibitem{lloyd1996universal}
Seth Lloyd.
\newblock Universal quantum simulators.
\newblock {\em Science}, 273(5278):1073--1078, 1996.

\bibitem{Blatt2012Quantum}
Rainer Blatt and Christian~F Roos.
\newblock Quantum simulations with trapped ions.
\newblock {\em Nat. Phys.}, 8(4):277--284, 2012.

\bibitem{Nielsen2010Quantum}
Michael~A Nielsen and Isaac~L Chuang.
\newblock {\em Quantum computation and quantum information: 10th anniversary
  edition}.
\newblock Cambridge University Press, 2010.

\bibitem{Monz2016Realization}
Thomas Monz, Daniel Nigg, Esteban~A Martinez, Matthias~F Brandl, Philipp
  Schindler, Richard Rines, Shannon~X Wang, Isaac~L Chuang, and Rainer Blatt.
\newblock Realization of a scalable shor algorithm.
\newblock {\em Science}, 351(6277):1068--1070, 2016.

\bibitem{peruzzo2014a}
Alberto Peruzzo, Jarrod Mcclean, Peter Shadbolt, Man-Hong Yung, Xiao-Qi Zhou,
  Peter~J Love, Al\'{a}n Aspuru-Guzik, and Jeremy~L O'brien.
\newblock A variational eigenvalue solver on a photonic quantum processor.
\newblock {\em Nat. Commun.}, 5(4213):4213--4213, 2014.

\bibitem{shen2017quantum}
Yangchao Shen, Xiang Zhang, Shuaining Zhang, Jing-Ning Zhang, Man-Hong Yung,
  and Kihwan Kim.
\newblock Quantum implementation of the unitary coupled cluster for simulating
  molecular electronic structure.
\newblock {\em Phys. Rev. A}, 95(2), 2017.

\bibitem{Ivanov2015Efficient}
Svetoslav~S Ivanov, Peter~A Ivanov, and Nikolay~V Vitanov.
\newblock Efficient construction of three- and four-qubit quantum gates by
  global entangling gates.
\newblock {\em Phys. Rev. A}, 91:032311, Mar 2015.

\bibitem{martinez2016compiling}
Esteban~A Martinez, Thomas Monz, Daniel Nigg, Philipp Schindler, and Rainer
  Blatt.
\newblock Compiling quantum algorithms for architectures with multi-qubit
  gates.
\newblock {\em New J. Phys.}, 18(6):063029, 2016.

\bibitem{Maslov2018Use}
Dmitri Maslov and Yunseong Nam.
\newblock Use of global interactions in efficient quantum circuit
  constructions.
\newblock {\em New J. Phys.}, 20(3):033018, 2018.

\bibitem{Casanova2012Quantum}
Jorge Casanova, Antonio Mezzacapo, Lucas Lamata, and Enrique Solano.
\newblock Quantum simulation of interacting fermion lattice models in trapped
  ions.
\newblock {\em Phys. Rev. Lett.}, 108:190502, May 2012.

\bibitem{yung2015from}
Man-Hong Yung, Jorge Casanova, Antonio Mezzacapo, Jarrod Mcclean, Lucas Lamata,
  Alan Aspuru-Guzik, and Enrique Solano.
\newblock From transistor to trapped-ion computers for quantum chemistry.
\newblock {\em Sci. Rep.}, 4(1):3589, 2015.

\bibitem{garciaalvarez2017digital}
Laura Garc\'{i}a-\'{A}lvarez, \'{I}\~{n}igo~Luis Egusquiza, Lucas Lamata,
  Adolfo~Del Campo, Julian Sonner, and Enrique Solano.
\newblock Digital quantum simulation of minimal ads/cft.
\newblock {\em Phys. Rev. Lett.}, 119(4):040501, 2017.

\bibitem{Hempel2018Quantum}
Cornelius Hempel, Christine Maier, Jonathan Romero, Jarrod McClean, Thomas
  Monz, Heng Shen, Petar Jurcevic, Ben~P Lanyon, Peter Love, Ryan Babbush,
  Al\'an Aspuru-Guzik, Rainer Blatt, and Christian~F Roos.
\newblock Quantum chemistry calculations on a trapped-ion quantum simulator.
\newblock {\em Phys. Rev. X}, 8:031022, Jul 2018.

\bibitem{Kim2009Entanglement}
Kihwan Kim, Ming-Shien Chang, Rajibul Islam, Simcha Korenblit, Lu-Ming Duan,
  and Christopher Monroe.
\newblock Entanglement and tunable spin-spin couplings between trapped ions
  using multiple transverse modes.
\newblock {\em Phys. Rev. Lett.}, 103(12):120502, 2009.

\bibitem{korenblit2012quantum}
Simcha Korenblit, Dvir Kafri, Wess~C Campbell, Rajibul Islam, Emily~E Edwards,
  Zhe-Xuan Gong, Guin-Dar Lin, Lu-Ming Duan, Jungsang Kim, Kihwan Kim, and
  Christopher Monroe.
\newblock Quantum simulation of spin models on an arbitrary lattice with
  trapped ions.
\newblock {\em New J. Phys.}, 14(9):095024--095024, 2012.

\bibitem{linke2017experimental}
Norbert~M Linke, Dmitri Maslov, Martin Roetteler, Shantanu Debnath, Caroline
  Figgatt, Kevin~A Landsman, Kenneth Wright, and Christopher Monroe.
\newblock Experimental comparison of two quantum computing architectures.
\newblock {\em Proc. Natl. Acad. Sci. U. S. A.}, 114(13):3305--3310, 2017.

\bibitem{molmer1999multiparticle}
Klaus M{\o}lmer and Anders S{\o}rensen.
\newblock Multiparticle entanglement of hot trapped ions.
\newblock {\em Phys. Rev. Lett.}, 82(9):1835--1838, 1999.

\bibitem{sorensen1999quantum}
Anders~S S{\o}rensen and Klaus M{\o}lmer.
\newblock Quantum computation with ions in thermal motion.
\newblock {\em Phys. Rev. Lett.}, 82(9):1971--1974, 1999.

\bibitem{solano1999deterministic}
Enrique Solano, Ruynet~Lima de~Matos~Filho, and Nicim Zagury.
\newblock Deterministic bell states and measurement of the motional state of
  two trapped ions.
\newblock {\em Phys. Rev. A}, 59(4):029903, 1999.

\bibitem{lanyon2011universal}
Ben~P Lanyon, Cornelius Hempel, Daniel Nigg, Markus M\"{u}ller, Rene Gerritsma,
  F~Z\"{a}hringer, Philipp Schindler, Julio~T Barreiro, Markus Rambach, Gerhard
  Kirchmair, Markus Hennrich, Peter Zoller, Rainer Blatt, and Christian~F Roos.
\newblock Universal digital quantum simulation with trapped ions.
\newblock {\em Science}, 334(6052):57--61, 2011.

\bibitem{barreiro2011an}
Julio~T Barreiro, Markus M\"{u}ller, Philipp Schindler, Daniel Nigg, Thomas
  Monz, Michael Chwalla, Markus Hennrich, Christian~F Roos, Peter Zoller, and
  Rainer Blatt.
\newblock An open-system quantum simulator with trapped ions.
\newblock {\em Nature}, 470(7335):486--491, 2011.

\bibitem{Monz201114}
Thomas Monz, Philipp Schindler, Julio~T Barreiro, Michael Chwalla, Daniel Nigg,
  William~A Coish, Max Harlander, Wolfgang H{\"a}nsel, Markus Hennrich, and
  Rainer Blatt.
\newblock 14-qubit entanglement: creation and coherence.
\newblock {\em Phys. Rev. Lett.}, 106(13):130506, 2011.

\bibitem{Zhu2006Arbitrary}
Shi-Liang Zhu, Christopher Monroe, and Lu-Ming Duan.
\newblock Arbitrary-speed quantum gates within large ion crystals through
  minimum control of laser beams.
\newblock {\em Europhys. Lett.}, 73(4):485--491, 2006.

\bibitem{zhu2006trapped}
Shi-Liang Zhu, Christopher Monroe, and Lu-Ming Duan.
\newblock Trapped ion quantum computation with transverse phonon modes.
\newblock {\em Phys. Rev. Lett.}, 97(5):050505, 2006.

\bibitem{Steane2014Pulsed}
Andrew~M Steane, G~Imreh, Jonathan~P Home, and Dietrich Leibfried.
\newblock Pulsed force sequences for fast phase-insensitive quantum gates in
  trapped ions.
\newblock {\em New J. Phys.}, 16(5):053049, 2014.

\bibitem{Green2014Phase}
Todd~J Green and Michael~J Biercuk.
\newblock Phase-modulated decoupling and error suppression in qubit-oscillator
  systems.
\newblock {\em Phys. Rev. Lett.}, 114(12):120502, 2014.

\bibitem{Leung2018Robust}
Pak~Hong Leung, Kevin~A Landsman, Caroline Figgatt, Norbert~M Linke,
  Christopher Monroe, and Kenneth~R Brown.
\newblock Robust 2-qubit gates in a linear ion crystal using a
  frequency-modulated driving force.
\newblock {\em Phys. Rev. Lett.}, 120(2):020501, 2018.

\bibitem{Choi2014Optimal}
Taeyoung Choi, Shantanu Debnath, Andrew~T Manning, Caroline Figgatt, Zhe-Xuan
  Gong, Lu-Ming Duan, and Christopher Monroe.
\newblock Optimal quantum control of multimode couplings between trapped ion
  qubits for scalable entanglement.
\newblock {\em Phys. Rev. Lett.}, 112(19):190502, 2014.

\bibitem{milne2018phase}
Alistair~R Milne, Claire~L Edmunds, Cornelius Hempel, Virginia Frey, Sandeep
  Mavadia, and Michael~J Biercuk.
\newblock Phase-modulated entangling gates robust against static and
  time-varying errors.
\newblock {\em arXiv:1808.10462}, 2018.

\bibitem{fisk1997accurate}
P~T~H Fisk, M~J Sellars, M~A Lawn, and G~Coles.
\newblock Accurate measurement of the 12.6 ghz "clock" transition in trapped
  /sup 171/yb/sup +/ ions.
\newblock {\em IEEE Trans. Ultrason. Ferroelectr. Freq. Control},
  44(2):344--354, 1997.

\bibitem{Olmschenk2007Manipulation}
Steve Olmschenk, Kelly~C Younge, David~L Moehring, Dzmitry~N Matsukevich, Peter
  Maunz, and Christopher Monroe.
\newblock Manipulation and detection of a trapped yb(+) hyperfine qubit.
\newblock {\em Phys. Rev. A}, 76(5):052314--052314, 2007.

\bibitem{hayes2010entanglement}
David Hayes, Dzmitry~N Matsukevich, Peter Maunz, David Hucul, Qudsia Quraishi,
  Steve Olmschenk, Wesley~C Campbell, Jonathan Mizrahi, Crystal Senko, and
  Christopher Monroe.
\newblock Entanglement of atomic qubits using an optical frequency comb.
\newblock {\em Phys. Rev. Lett.}, 104(14):140501, 2010.

\bibitem{Debnath2016Demonstration}
Shantanu Debnath, Norbert~M Linke, Caroline Figgatt, Kevin~A Landsman, Ken
  Wright, and Christopher Monroe.
\newblock Demonstration of a small programmable quantum computer with atomic
  qubits.
\newblock {\em Nature}, 536(7614):63, 2016.

\bibitem{haljan2005spin-dependent}
Paul~C Haljan, Kathy-Anne Brickman, Louis Deslauriers, Patricia~J Lee, and
  Christopher Monroe.
\newblock Spin-dependent forces on trapped ions for phase-stable quantum gates
  and entangled states of spin and motion.
\newblock {\em Phys. Rev. Lett.}, 94(15):153602, 2005.

\bibitem{Lee2005Phase}
Patricia~J Lee, Kathy-Anne Brickman, Louis Deslauriers, Paul~C Haljan, Lu-Ming
  Duan, and Christopher Monroe.
\newblock Phase control of trapped ion quantum gates.
\newblock {\em J. Opt. B}, 7(10):S371--S383, 2005.

\bibitem{James1998Quantum}
Daniel F.~V. James.
\newblock Quantum dynamics of cold trapped ions with application to quantum
  computation.
\newblock {\em Appl. Phys. B: Lasers Opt.}, 66(2):181--190, 1998.

\bibitem{hayes2012coherent}
David Hayes, Susan~M Clark, Shantanu Debnath, David Hucul, I~Volkan Inlek,
  Kenny~W Lee, Qudsia Quraishi, and Christopher Monroe.
\newblock Coherent error suppression in multiqubit entangling gates.
\newblock {\em Phys. Rev. Lett.}, 109(2):020503, 2012.

\bibitem{webb2018resilient}
Anna~E Webb, Simon~C Webster, S.~Collingbourne, David Bretaud, Adam~M Lawrence,
  Sebastian Weidt, Florian Mintert, and Winfried~K Hensinger.
\newblock Resilient entangling gates for trapped ions.
\newblock {\em Phys. Rev. Lett.}, 121:180501, Nov 2018.

\bibitem{shapira2018robust}
Yotam Shapira, Ravid Shaniv, Tom Manovitz, Nitzan Akerman, and Roee Ozeri.
\newblock Robust entanglement gates for trapped-ion qubits.
\newblock {\em Phys. Rev. Lett.}, 121:180502, Nov 2018.

\bibitem{Roos2008Ion}
Christian~F Roos.
\newblock Ion trap quantum gates with amplitude-modulated laser beams.
\newblock {\em New J. Phys.}, 10(1):013002, 2008.

\bibitem{sackett2000experimental}
C~A Sackett, David Kielpinski, B~E King, Christopher Langer, V~Meyer, C~J
  Myatt, M~Rowe, Q~A Turchette, Wayne~M Itano, David~J Wineland, and
  Christopher Monroe.
\newblock Experimental entanglement of four particles.
\newblock {\em Nature}, 404(6775):256--259, 2000.

\bibitem{Duan2012Correcting}
Lu-Ming Duan and Chao Shen.
\newblock Correcting detection errors in quantum state engineering through data
  processing.
\newblock {\em New J. Phys.}, 14(5):1778--1782, 2012.

\bibitem{figgatt2018parallel}
Caroline Figgatt, Aaron Ostrander, Norbert~M Linke, Kevin~A Landsman, Daiwei
  Zhu, Dmitri Maslov, and Christopher Monroe.
\newblock Parallel entangling operations on a universal ion trap quantum
  computer.
\newblock {\em arXiv:1810.11948}, 2018.

\end{thebibliography}

\clearpage
\onecolumngrid
\section{Methods}

\subsection{Expressions of Constraints under Discrete Phase Modulation}
Here, we give the detailed expressions of the constraints under the discrete phase modulation. Review the constraints shown in main text,
\begin{align}
  \alpha_{j,m}(\tau) & = - \ii \eta_{j,m}
        \int_{0}^{\tau}
        \dfrac{\Omega_{j}(t)e^{-\ii \phi_{j}(t)}}{2}
        e^{\ii \left( \nu_m - \mu \right) t}
        \diff{t} = 0,  \label{equ:trajectory_SI}\\
  \theta_{j,j'}(\tau) &
    = -\sum_{m}
    \int_{0}^{\tau} \diff{t_2}
    \int_{0}^{t_2} \diff{t_1}
    \dfrac{\eta_{j,m} \eta_{j',m} \Omega_{j}(t_2) \Omega_{j'}(t_1)}{2}
  \sin\left[ (\nu_m - \mu)(t_2 - t_1) - \left( \phi_j(t_2) - \phi_{j'}(t_1) \right) \right] = \dfrac{\pi}{4}, \label{equ:coupling_SI}
\end{align}
where $\alpha_{j,m}(\tau)$ is the residual displacement of the $j$-th qubit and $m$-th motional mode at the gate time, and $\theta_{j,j'}(\tau)$ is the coupling strength between the $j$-th and the $j'$-th qubits. In the experiment we fix the total gate time to be $\tau$ and divide it into $K$ segments with the segment duration of $\tau_\mathrm{s} = \tau / K$. The phases are modulated discretely with the form of
\begin{equation}\label{equ:phase_SI}
  \phi_j(t) = \left\{
  \begin{aligned}
  & \phi_{j,1} & 0 \leq t \leq \tau_\mathrm{s} \\
  & ... \\
  & \phi_{j,k} & (k-1)\tau_\mathrm{s} \leq t \leq k\tau_\mathrm{s} \\
  & ... \\
  & \phi_{j,K} & \tau - \tau_\mathrm{s} \leq t \leq \tau \\
  \end{aligned}
  \right. ,
\end{equation}
where $\phi_{j,k}$ is the phase of the Rabi frequency on the $j$-th qubit in the $k$-th segment. And the time dependent amplitude of Rabi frequency can be written as
\begin{equation}\label{equ:amp_SI}
  \Omega_j(t) = \Omega_{j}^{\mathrm{max}} w(t),
\end{equation}
where $\Omega_{j}^{\mathrm{max}}$ is the maximal value of the amplitude applied on the $j$-th qubit and $w(t)$ is the pulse-shaping function to slowly turn on (off) amplitude at the first (last) segment with the form of $\sin^2-$profile
\begin{equation}\label{equ:wf_SI}
  w(t) = \left\{
  \begin{aligned}
    & \sin^2\left(\dfrac{\pi}{2\tau_\mathrm{s}}t\right) & 0\leq t \leq\tau_s \\
    & 1 & \tau_{\mathrm{s}}\leq t \leq\tau-\tau_{\mathrm{s}}\\
    &  \sin^2\left(\dfrac{\pi}{2\tau_s}\left(t - \tau\right)\right) & \tau-\tau_s\leq t \leq\tau
  \end{aligned}
  \right. .
\end{equation}
By inserting the pulse scheme of \equref{equ:phase_SI} and \equref{equ:wf_SI} into \equref{equ:trajectory_SI}, we can rewrite the residual displacements as
\begin{align}
  \alpha_{j,m}(\tau) & = \eta_{j,m}\Omega_{j}^{\mathrm{max}} d_{j,m}/2 \\
  d_{j,m} & = \sum_{k = 1}^{K}
    \left(
        \mathrm{Ts}_{m,k} \mathrm{X}_{j,k} + \mathrm{Tc}_{m,k} \mathrm{Y}_{j,k}
    \right)
    + \ii
    \sum_{k = 1}^{K}
    \left(
        \mathrm{Tc}_{m,k} \mathrm{X}_{j,k} - \mathrm{Ts}_{m,k} \mathrm{Y}_{j,k}
    \right) \label{equ:scaled_traj_SI},
\end{align}
where $d_{j,m}$ is the scaled residual displacement of $\alpha_{j,m}(\tau)$ and
\begin{align}
  \mathrm{Ts}_{m,k} & =
    \int_{(k-1)\tau_\mathrm{s}}^{k\tau_\mathrm{s}}
    w(t) \sin\left[ \left( \nu_m - \mu \right)t \right] ,\\
  \mathrm{Tc}_{m,k} & =
    -\int_{(k-1)\tau_\mathrm{s}}^{k\tau_\mathrm{s}}
    w(t) \cos\left[ \left( \nu_m - \mu \right)t \right]   ,\\
  \mathrm{X}_{j,k} & =
    \cos \phi_{j,k} ,\\
  \mathrm{Y}_{j,k} & =
    \sin \phi_{j,k} .
\end{align}
Similarly, for the coupling strength of \equref{equ:coupling_SI} we can also rewrite it as
\begin{align}
  \theta_{j,j'}(\tau) &= \Omega_{j}^{\mathrm{max}}\Omega_{j'}^{\mathrm{max}}g_{j,j'}, \\
  g_{j,j'} & =  \sum_{k,l}
        \mathrm{X}_{j,k} \mathrm{Gs}_{j,j',k,l} \mathrm{X}_{j',l}
        + \mathrm{Y}_{j,k} \mathrm{Gs}_{j,j',k,l} \mathrm{Y}_{j',l}
        + \mathrm{X}_{j,k} \mathrm{Gc}_{j,j',k,l} \mathrm{Y}_{j',l}
        - \mathrm{Y}_{j,k} \mathrm{Gc}_{j,j',k,l} \mathrm{X}_{j',l}
         \label{equ:scaled_coupling_SI},
\end{align}
where $g_{j,j'}$ is the rescaled coupling strength of $\theta_{j,j'}(\tau)$ and
\begin{equation}
  \mathrm{Gs}_{j,j',k,l} =
  \left\{
  \begin{aligned}
    & -\sum_m
    \dfrac{\eta_{j,m}\eta_{j',m}}{2}
    \int_{(k-1)\tau_\mathrm{s}}^{k\tau_\mathrm{s}}
    \int_{(l-1)\tau_\mathrm{s}}^{l\tau_\mathrm{s}}
    \diff{t_2}\diff{t_1}
    w(t_2)w(t_1)
    \sin\left[ (\nu_m - \mu)(t_2 - t_1) \right]
    & \quad l < k \\
    & -\sum_m
    \dfrac{\eta_{j,m}\eta_{j',m}}{2}
    \int_{(k-1)\tau_\mathrm{s}}^{k\tau_\mathrm{s}}
    \int_{(k-1)\tau_\mathrm{s}}^{t_2}
    \diff{t_2}\diff{t_1}
    w(t_2)w(t_1)
    \sin\left[ (\nu_m - \mu)(t_2 - t_1) \right]
    & \quad l = k \\
    & 0 & \quad l>k
  \end{aligned}
  \right. ,
\end{equation}
\begin{equation}
  \mathrm{Gc}_{j,j',k,l} =
  \left\{
  \begin{aligned}
    & -\sum_m
     \dfrac{\eta_{j,m}\eta_{j',m}}{2}
    \int_{(k-1)\tau_\mathrm{s}}^{k\tau_\mathrm{s}}
    \int_{(l-1)\tau_\mathrm{s}}^{l\tau_\mathrm{s}}
    \diff{t_2}\diff{t_1}
    w(t_2)w(t_1)
    \cos\left[ (\nu_m - \mu)(t_2 - t_1) \right]
    & \quad l < k \\
    & -\sum_m
     \dfrac{\eta_{j,m}\eta_{j',m}}{2}
    \int_{(k-1)\tau_\mathrm{s}}^{k\tau_\mathrm{s}}
    \int_{(k-1)\tau_\mathrm{s}}^{t_2}
    \diff{t_2}\diff{t_1}
    w(t_2)w(t_1)
    \cos\left[ (\nu_m - \mu)(t_2 - t_1) \right]
    & \quad l = k \\
    & 0 & \quad l>k
  \end{aligned}
  \right. .
\end{equation}
It is convenient to rearrange \equref{equ:scaled_traj_SI} and \equref{equ:scaled_coupling_SI} into the matrix-form of
\begin{align}\label{equ:matrix_SI}
  d_{j,m}&=
    \left(
      \mathbf{Ts}_m^{\mathrm{T}}\mathbf{X}_j + \mathbf{Tc}_m^{\mathrm{T}}\mathbf{Y}_j
    \right)
    +\ii
    \left(
      \mathbf{Tc}_m^{\mathrm{T}}\mathbf{X}_j - \mathbf{Ts}_m^{\mathrm{T}}\mathbf{Y}_j
    \right) , \\
  g_{j,j'}&=
    \mathbf{X}_j^\mathrm{T} \mathbf{Gs}_{j,j'} \mathbf{X}_{j'} + \mathbf{Y}_{j}^\mathrm{T} \mathbf{Gs}_{j,j'} \mathbf{Y}_{j'}+ \mathbf{X}_{j}^\mathrm{T} \mathbf{Gc}_{j,j'} \mathbf{Y}_{j'}           - \mathbf{Y}_{j}^\mathrm{T} \mathbf{Gc}_{j,j'} \mathbf{X}_{j'}.
\end{align}
Here $\mathbf{Ts}_{m}$, $\mathbf{Tc}_{m}$, $\mathbf{X}_{j}$ and $\mathbf{Y}_{j}$ are the column vectors of $\left\{ \mathrm{Ts}_{m,k} \right\}$, $\left\{ \mathrm{Tc}_{m,k} \right\}$, $\left\{ \mathrm{X}_{j,k} \right\}$ and $\left\{ \mathrm{Y}_{j,k} \right\}$ respectively, while $\mathbf{Gs}_{j,j'}$ and $\mathbf{Gc}_{j,j'}$ are the matrix-form of $\left\{ \mathrm{Gs}_{j,j',k,l} \right\}$ and $\left\{ \mathrm{Gc}_{j,j',k,l} \right\}$ respectively.

Based on the above calculations, we summarize the constraints as below,
\begin{align}
  &d_{j,m}= 0, \label{equ:matrix_traj_SI}\\
  &\Omega_{j}^{\mathrm{max}}\Omega_{j'}^{\mathrm{max}} g_{j,j'} = \pi/4, \label{equ:matrix_coup_SI}\\
  &\mathbf{X}_{j}^{\mathrm{T}}\mathbf{X}_{j} + \mathbf{Y}_{j}^{\mathrm{T}}\mathbf{Y}_{j} = \mathbf{1}, \label{equ:matrix_phase_SI}
\end{align}
for all $j,j',m$.

\subsection{Pulse Scheme Optimization}
According to the \equref{equ:matrix_SI} to \equref{equ:matrix_phase_SI}, although we have already written all the constraints in the matrix-form, it is challenging to directly solve the equations of the constraints due to nonlinearity of \equref{equ:matrix_coup_SI} and \equref{equ:matrix_phase_SI}. Instead we construct an optimization problem by minimizing the objective function,
\begin{equation}\label{equ:objectivef_SI}
  \sum_{j,m}\left| d_{j,m} \right|^2,
\end{equation}
which is equivalent to $\sum_{j,m}\left| \alpha_{j,m}(\tau) \right|^2$, subject to the constraints of \equref{equ:matrix_coup_SI} and \equref{equ:matrix_phase_SI}. The construction is still non-trivial because we want to efficiently obtain the suitable pulse scheme. The main difficulty in performing optimization is to fulfill the non-linear constraints and moreover, the number of non-linear constraints grows quadratically with the number of qubits. To simplify the optimization problem, we utilize the symmetries of the Lamb-Dicke parameters, which always have the relations of $\eta_{j,m} = \pm \eta_{N-j+1,m}$, then set the $\phi_{j}(t)$ and $\Omega_{j}^{\mathrm{max}}$ to be same for the ions $j$ and $(N-j+1)$. Taking the four-ion case as an example, the constraints of the coupling strengths are reduced from
\begin{equation}
  \theta_{1,2}(\tau)=
  \theta_{1,3}(\tau)=
  \theta_{1,4}(\tau)=
  \theta_{2,3}(\tau)=
  \theta_{2,4}(\tau)=
  \theta_{3,4}(\tau)=
  \pi/4,
\end{equation}
to
\begin{equation}\label{equ:4qubitconstraints_SI}
  \theta_{1,2}(\tau)=
  \theta_{1,3}(\tau)=
  \theta_{1,4}(\tau)=
  \theta_{2,3}(\tau)=
  \pi/4,
\end{equation}
because $\theta_{1,2}(\tau) = \theta_{3,4}(\tau)$ and $\theta_{1,3}(\tau) = \theta_{2,4}(\tau)$ always establish. If we rewrite \equref{equ:4qubitconstraints_SI} with the scaled coupling strength and utilize the relations of $\Omega_{1}^{\mathrm{max}} = \Omega_{4}^{\mathrm{max}}$ and $\Omega_{2}^{\mathrm{max}} = \Omega_{3}^{\mathrm{max}}$, we further simplify the non-linear constraints to
\begin{align}\label{equ:finalconstraints_SI}
  \nonumber
  g_{1,2} & = g_{1,3}\\
  g_{1,2}* g_{1,3} & = g_{1,4}* g_{2,3}.
\end{align}
Finally we construct the optimization problem of minimizing the objective function of $\sum_{j=1}^{2}\sum_{m=1}^{4}\left| d_{j,m} \right|^2$ subject to the constraints of \equref{equ:finalconstraints_SI} and \equref{equ:matrix_phase_SI}. After obtaining the modulated phase patterns we solve the \equref{equ:matrix_coup_SI} to get the theoretical values of the maximal amplitudes of the Rabi frequencies $\left\{ \Omega_j^{\mathrm{max}} \right\}$. Moreover,
we manually introduce an additional symmetry to the modulated patterns by presetting the modulated phases to be $\phi_{j}(t) = -\phi_{j}(\tau - t)$ or $\phi_{j,k} = -\phi_{j, K-k+1}$ before the optimizing procedure.

\subsection{Experimental Setup}
\begin{figure}[htb]
  \centering
  \includegraphics[scale = 0.9]{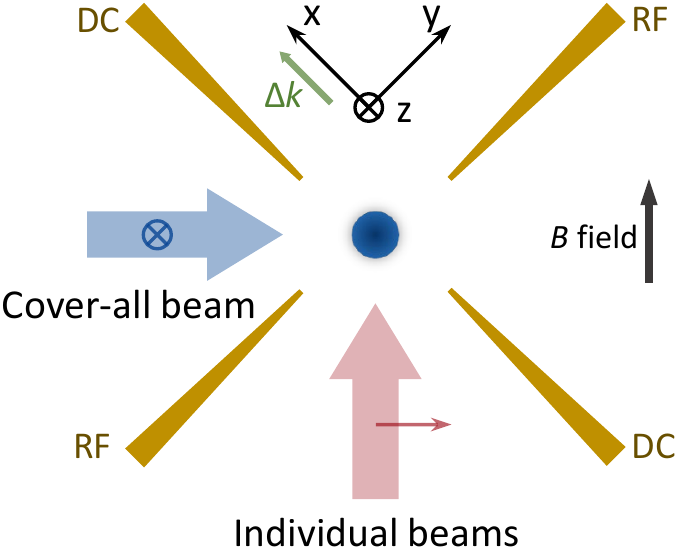}
  \caption{
  {\bf Side view of the experimental ion-trap system.} The direction of the magnetic {\it{B}} field is shown in the figure and the value of it is around 6 Gauss.
  The cover-all beam goes through the side-viewport, while the individual beams go through the bottom re-entrant viewport to achieve the focused waist of $\sim 1~\mu \mathrm{m}$.
  The effective wave-vector of the two Raman beams is almost in the x-direction,
  and the polarizations of them are both linear and perpendicular to each other.
  }\label{fig:setup_SI}
\end{figure}

In the experiment the single ion-chain is held in a blade trap, with the geometry shown in \figref{fig:setup_SI}. The Raman beams are produced by a pico-second pulse laser with the center wavelength of $377~\mathrm{nm}$ and the repetition rate of $\sim 76\mathrm{MHz}$. The ions fluorescence during the detection process is collected by the objective lens from the top re-entrant viewport then imaged to the EMCCD.

\subsection{Experimental Parameters}
Here we present the details of the experimental pulse schemes for the global 3-qubit and 4-qubit entangling gates. The maximal amplitudes of the Rabi frequencies are given based on the theoretical Lamb-Dicke parameters,
\begin{equation}\label{equ_LDP_SI}
  \eta_{j,m} =
        b_{j,m}
        \dfrac{2\sqrt{2}\pi}{\lambda}
        \sqrt{
            \dfrac{\hbar}{2M_{\mathrm{Yb}}\nu_m}
        },
\end{equation}
where $b_{j,m}$ is the element of the normal mode transformation matrix for the ion $j$ and the motional mode $m$, $\lambda$ is the center wavelength of the Raman laser, $\hbar$ is the reduced Planck constant and $M_{\mathrm{Yb}}$ is the mass of the $\yb$ ion. 
The specific values of the modulated phases and amplitudes of Rabi frequencies obtained through the optimization are shown in \tabref{tab:3qubit_SI} and \tabref{tab:4qubit_SI}. In the experimental realization the required amplitudes of the Rabi frequencies are larger than the theoretical calculation due to the overestimation of the Lamb-Dicke parameters.
\renewcommand\arraystretch{1.3}
\begin{table}
  \centering
  \caption{Pulse scheme for the global 3-qubit entangling gate}\label{tab:3qubit_SI}
  \begin{tabular}{c|c|c|c|c}
    \toprule
    \multicolumn{2}{c|}{qubit $j$} & 1 & 2 & 3 \\
    \cline{1-5}
    \multicolumn{2}{c|}{$\Omega_j^{\mathrm{max}}$ (MHz)} & $-2\pi \times 0.181$ & $2\pi \times 0.253$ & $-2\pi \times 0.181$ \\
    \cline{1-5}
    \multirow{6}*{$\phi_{j,k} (\pi)$}& 1 & 0.104 & 0.104 & 0.104 \\
    & 2 & 0.033 & 0.033 & 0.033 \\
    & 3 & 0.095 & 0.095 & 0.095 \\
    & 4 & -0.095 & -0.095 & -0.095 \\
    & 5 & -0.033 & -0.033 & -0.033 \\
    & 6 & -0.104 & -0.104 & -0.104 \\
    \toprule
  \end{tabular}
\end{table}

\begin{table}
  \centering
  \caption{Pulse scheme for the global 4-qubit entangling gate}\label{tab:4qubit_SI}
  \begin{tabular}{c|c|c|c|c|c}
    \toprule
    \multicolumn{2}{c|}{qubit $j$} & 1 & 2 & 3  & 4\\
    \cline{1-6}
    \multicolumn{2}{c|}{$\Omega_j^{\mathrm{max}}$ (MHz)} & $-2\pi \times 0.117$ & $2\pi \times 0.168$ & $2\pi \times 0.168$ & $-2\pi \times 0.117$ \\
    \cline{1-6}
    \multirow{12}*{$\phi_{j,k} (\pi)$}& 1 & 0.041 & 0.231 & 0.231 & 0.041 \\
    & 2 & -0.070 & 0.579 & 0.579 & -0.070 \\
    & 3 & 0.472 & -0.001 & -0.001 & 0.472 \\
    & 4 & 0.054 & 0.230 & 0.230 & 0.054 \\
    & 5 & 0.035 & 0.285 & 0.285 & 0.035 \\
    & 6 & 0.402 & -0.170 & -0.170 & 0.402\\
    & 7 & -0.402 & 0.170 & 0.170 & -0.402\\
    & 8 & -0.035 & -0.285 & -0.285 & -0.035 \\
    & 9 & -0.054 & -0.230 & -0.230 & -0.054 \\
    & 10 & -0.472 & 0.001 & 0.001 & -0.472 \\
    & 11 & 0.070 & -0.579 & -0.579 & 0.070 \\
    & 12 & -0.041 & -0.231 & -0.231 & -0.041 \\
    \toprule
  \end{tabular}
\end{table}

\begin{figure}[htb]
  \centering
  \includegraphics[scale = 0.6]{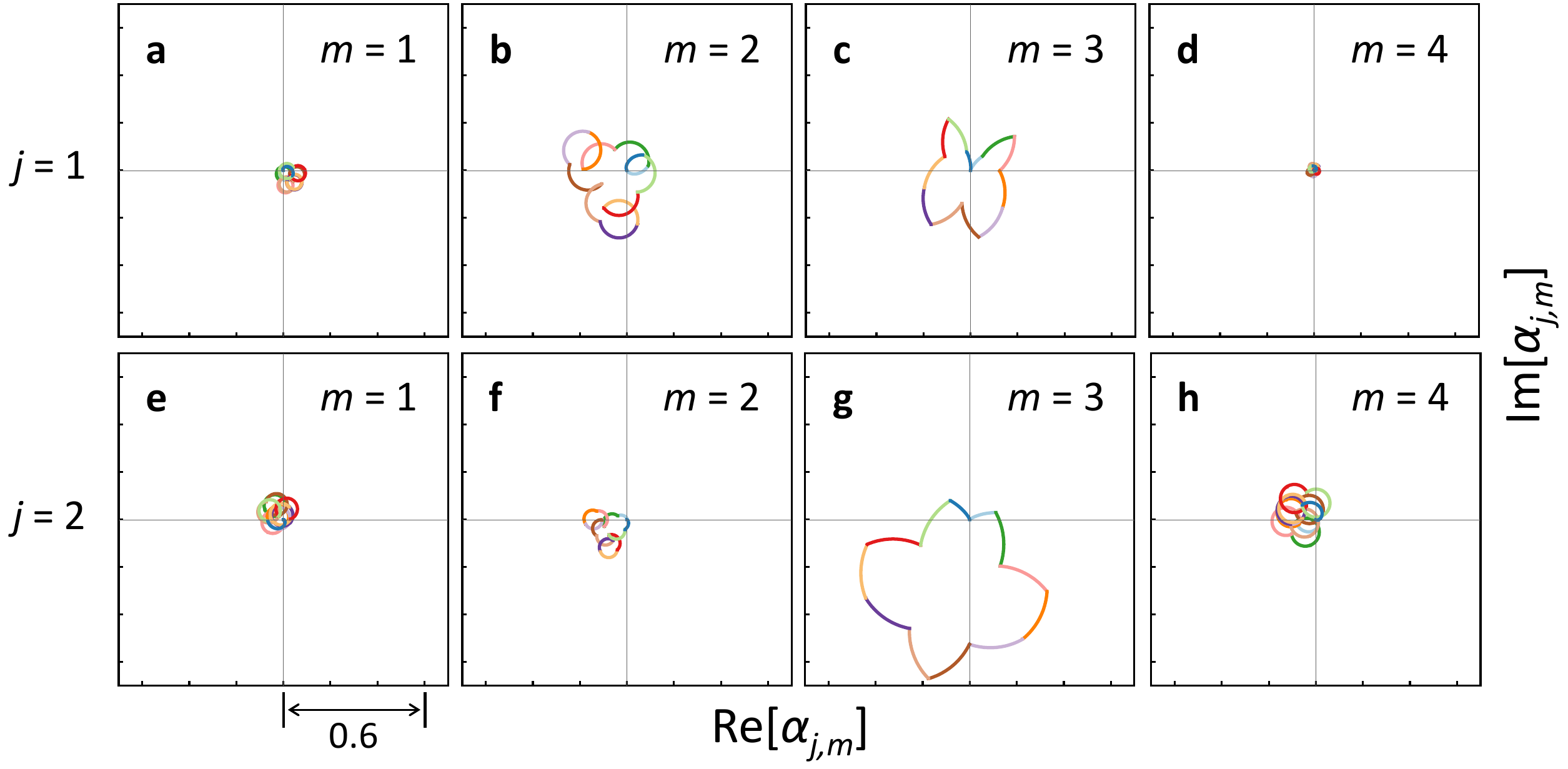}
  \caption{
  {\bf Motional trajectories in the phase space for the global 4-qubit entangling gate.}
  Because we apply different modulated phase patterns to the qubits $\left\{ 1,4 \right\}$ and the qubits $\left\{ 2,3 \right\}$, the shapes of the motional trajectories are different, as shown in {\bf a}-{\bf d} and {\bf e}-{\bf h}, respectively.
  }\label{fig:4qubits_traj_SI}
\end{figure}

In the main text we have already shown the trajectories of the motional modes in the phase space for the 3-qubit situation. Here, we supplement the motion trajectories of $\alpha_{j,m}(t)$ for the 4-qubit situation, as shown in \figref{fig:4qubits_traj_SI}.

\end{document}